\documentclass[12pt,english,floatfix,nofootinbib,superscriptaddress,aps,prd,preprint]{revtex4}
\usepackage[utf8]{inputenc}
\usepackage{float}
\usepackage{array}
\usepackage{bbold}
\usepackage{lipsum}
\usepackage{dsfont}
\usepackage{graphicx}
\usepackage{amsmath,amsthm,amsfonts,amssymb}
\usepackage{graphicx}
\usepackage[english]{babel} 
\usepackage{color}
\usepackage{tensor}
\usepackage{esint}
\usepackage[dvips]{epsfig}
\usepackage[dvips]{graphicx}
\usepackage{float}
\usepackage{units}
\usepackage{textcomp}
\usepackage{mathrsfs}
\usepackage{amsmath}
\usepackage[makeroom]{cancel}
\usepackage{amssymb}
\usepackage{amsbsy}
\usepackage{amsfonts}
\usepackage{amssymb,mathrsfs,xcolor}
\usepackage{esint}
\usepackage{braket}
\usepackage{array}
\usepackage{graphicx}

\usepackage{wasysym}
\usepackage{multirow}
\usepackage{wrapfig}
\usepackage{subfig}

\usepackage{stmaryrd}
\usepackage{upgreek}

\makeatletter

\makeatletter\usepackage{babel}

\usepackage{hyperref}
\hypersetup{
    colorlinks,
    citecolor=blue,
    filecolor=green,
    linkcolor=purple,
    urlcolor=red,
}

\usepackage{slashed}

\newcommand{\ie}{\begin{equation}}
\newcommand{\fe}{\end{equation}}
\newcommand{\se}{\begin{eqnarray}}
\newcommand{\ff}{\end{eqnarray}}

\begin{document}

\title{Fermions with Electric Dipole Moment in curved spacetime}

\author{A. A. Ara\'{u}jo Filho}
\email{dilto@fisica.ufc.br}

\affiliation{Departamento de Física Teórica and IFIC, Centro Mixto Universidad de Valencia--CSIC. Universidad
de Valencia, Burjassot-46100, Valencia, Spain}

\affiliation{Departamento de Física, Universidade Federal da Paraíba, Caixa Postal 5008, 58051--970, João Pessoa, Paraíba,  Brazil.}

\author{H. Hassanabadi}
\email{hha1349@gmail.com}

\affiliation{Physics Department, Shahrood University of Technology, Shahrood, Iran}

\author{J. A. A. S. Reis}
\email{jalfieres@gmail.com}

\affiliation{Universidade Estadual do Sudoeste da Bahia (UESB), Departamento de Ciências Exatas e Naturais, Campus Juvino Oliveira, Itapetinga -- BA, 45700-00,--Brazil}
 
\author{L. Lisboa-Santos}
\email{leticia.lisboa@discente.ufma.br} 

\affiliation{Programa de Pós-graduação em Física,  Universidade Federal do Cear\'a Campus do Pici, Fortaleza - CE, 60455-760, Brazil.}


\date{\today}

\begin{abstract}

This paper explores the relativistic behavior of spin--half particles possessing an Electric Dipole Moment (EDM) in a curved spacetime background induced by a spiral dislocation. A thorough review of the mathematical formulation of the Dirac spinor in the framework of quantum field theory sets the foundation for our investigation. By deriving the action that governs the interaction between the spinor field, the background spacetime, and an external electric field, we establish a framework to study the dynamics of the system. Solving the resulting wave equation reveals a set of coupled equations for the radial components of the Dirac spinor, which give rise to a modified energy spectrum attributed to the EDM. To validate our findings, we apply them to the geometric phase and thermodynamics.

\end{abstract}

\maketitle


\section{Introduction}

The Dirac equation \cite{1,thaller2013dirac} has been extensively explored across a wide range of scientific disciplines, demonstrating its significant relevance and broad applicability. Its investigations encompass various topics, including the Dirac oscillator \cite{2,3}, charged particles in static electromagnetic fields \cite{4}, pseudospin symmetry of nucleons in relativistic Manning--Rosen potentials \cite{5}, the behavior of Dirac particles in curved spacetimes \cite{6,7,obukhov2017general,8}, the Lorentz-- and CPT--violating standard model extension \cite{9,10,11,12,13,14}, electronic properties \cite{15,16}, topological defects of graphene \cite{18}, quantum rings \cite{17}, cosmic string spacetimes \cite{19}, and others \cite{20,21,22,23,24,25,26}. These comprehensive investigations collectively contribute to the enhanced understanding of the implications and applications of the Dirac equation in modern physics.

On the other hand, the influence of topology on physical systems has been widely studied across various domains of physics, including gravitation and condensed matter physics. The investigation of linear topological defects in crystalline solids and their connection to cosmology has received much attention over the past years \cite{27,28,29,30,31}. Some details of topological defects can be seen in Refs.\cite{volovik2001superfluid,volovik2003universe}. This intriguing approach has sparked numerous subsequent studies, addressing a range of topics. For instance, research has focused on the structure of the curvature tensor of cosmic strings associated with singularities \cite{32}, Kaluza--Klein theories \cite{37,38}, Klein--Gordon \cite{33,34}, spin--zero Duffin--Kemmer--Petiau \cite{39,40} and spin and pseudo--spin symmetries of Dirac particles \cite{41,42}.

Moreover, extensive investigations have explored the correlation between scalar fields and Coulomb--type potentials within spacetimes characterized by screw dislocations \cite{43}. Additionally, significant attention has been devoted to analyzing the non--relativistic quantum dynamics of particles harmonically interacting with conical singularities connected to cosmic strings, and global monopoles \cite{44}.
In Ref. \cite{silenko2007equivalence}, the authors reanalyzed the results of spin experiments with Hg atoms and established the first experimental bound on their anomalous gravitomagnetic moment. They suggested extending some experiments with spinning particles to test for the absence of the anomalous gravitomagnetic moment and to observe related gravitational effects.

Electrodynamics in noninertial frames is studied in \cite{obukhov2021electrodynamics}. Additionally, \cite{obukhov2016manifestations2} demonstrated that the inertia due to the Earth's rotation and gravitational fields affects the motion of elementary particles and their spin dynamics. This theme continues in \cite{obukhov2017general}, which investigated both classical and quantum mechanical theories of spinning particles with electric and magnetic dipole moments moving in arbitrary electromagnetic, inertial, and gravitational fields.

General relativity effects in precision spin experimental tests of fundamental symmetries were explored in \cite{vergeles2023general}. The influence of the Earth's rotation and gravity on particle motion and spin evolution was examined further in \cite{obukhov2016manifestations1}. Lastly, \cite{nikolaev2020maxwell} presented solutions to the Maxwell equations for electrostatic systems with manifestly vanishing electric currents in curved spacetime for stationary metrics, revealing a non--vanishing magnetic field of purely geometric origin. 
 
In this work, we consider the Dirac equation in a spiral spacetime with a topological defect. First, we determine the tetrads and their inverse, spin connection, spinorial connection, and the generalized Dirac metric. We then write the Dirac equation in the presence of the topological effect and examine the behavior of a fermion with an electric dipole moment interacting with an external electric field. By solving the Dirac equation, we determine the eigenvalues and components of the wave function, investigating the effects of the spiral dislocation parameters. Next, we consider a system including \( N \) identical such particles, determine the grand canonical ensemble, and obtain the grand potential of the system, including particle number, internal energy, entropy, and heat capacity. Finally, using the obtained internal energy, permittivity, and susceptibility of the system, we investigate the effect of the deflection on these parameters.


\section{Quantum dynamics in the presence of a spiral dislocation}

In light of the extensive literature on non--relativistic and relativistic quantum dynamics in spacetime involving dislocation fields as deformation fields, our objective is to provide a quantum description of a neutral spin--half particle with a PMDM (Proper Mass Distribution Matrix) interacting with an external electric field configuration within the relativistic regime. Specifically, we aim to investigate this interaction within a background spacetime generated by a distortion of a circle into a recognized spiral dislocation. In this particular case, such a field arises from a parameter associated with the spiral dislocation, denoted as $\chi$, which possesses the dimension of length \cite{45}. 

As previously noted by Bakke and Furtado \cite{46}, $\chi$ is directly related to the associated Burgers vector, denoted as $\mathbf{b}$, via the equation $\chi =\left\vert \mathbf{b} \right\vert /2\pi$. Referring to the discussion in Ref. \cite{45}, it is established that this vector is parallel to the plane $z=$ constant and directed radially. This type of linear topological defect, which is commonly described in the context of crystalline solids, serves as an inspiration for our study. Additionally, considering the connection between defects in solids and defects in cosmology, we generalize this linear defect to the realm of gravitation \cite{46,47,48}. 

Adopting natural units where $c=1$ and $\hbar =1$, the background spacetime resulting from this type of topological defect can be described by the following line element \cite{41}
\begin{equation}
\mathrm{d}s^{2}=\mathrm{d}t^{2}-\mathrm{d}r^{2}-2\chi \mathrm{d}r \mathrm{d}\varphi -\left( \chi ^{2}+r^{2}\right)
\mathrm{d}\varphi ^{2}-\mathrm{d}z^{2}.  \label{eq:2.1}
\end{equation}%

Here, the spatial component of the line element, as described by Eq. $\left(\ref{eq:2.1}\right)$, exhibits an intriguing distortion where a circle transforms into a spiral configuration known as the spiral dislocation. This distortion occurs within a plane perpendicular to the $z$ axis. It is important to note that we employ cylindrical coordinates, denoted as $r$, $\varphi$, and $z$, which represent the radial, azimuthal, and axial coordinates, respectively.

When studying the behavior of a neutral Dirac particle within a spacetime containing such a topological defect, it becomes necessary to examine Dirac spinors within a curved spacetime framework \cite{49}. Thereby, to define the spinors in the local reference frame of an observer, it is essential to establish a non--coordinate basis denoted by $\hat{\theta}^{a}$
\begin{equation}
\hat{\theta}^{a}=e_{\left. {}\right. \mu }^{a}\left( x\right) dx^{\mu }.
\label{eq:2.2}
\end{equation}%

The Minkowski spacetime metric tensor is represented by $\eta _{ab}=\text{diag}(+,-,-,-)$, where the Latin indices $a,b=0,1,2,3$ correspond to the local reference frame (LRF). On the other hand, the spacetime indices are denoted by $\mu$ and take values from the set $\{t,r,\varphi,z\}$. The inverse tetrads are determined by the equation $dx^{\mu}=e_{\left. {}\right. a}^{\mu }\left( x\right) 
\hat{\theta}^{a}$. Consequently, the relationships $e_{\left. {}\right. a}^{\mu
}\left( x\right) e_{\left. {}\right. \nu }^{a}\left( x\right) =\delta
_{\left. {}\right. \nu }^{\mu }$ and $e_{\left. {}\right. \mu }^{a}\left(
x\right) e_{\left. {}\right. b}^{\mu }\left( x\right) =\delta _{\left.
{}\right. b}^{a}$ are expected to be satisfied through the tetrads and their inverses. For convenience, the values of the inverse tetrads are presented as follows:
\begin{equation}
e_{\left. {}\right. 0}^{t}=e_{\left. {}\right. 1}^{r}=e_{\left. {}\right.
3}^{z}=1,\left. {}\right. e_{\left. {}\right. 2}^{\varphi }=1/r,\left.
{}\right. e_{\left. {}\right. 2}^{r}=-\chi /r.  \label{eq:2.4}
\end{equation}%

In our investigation of the behavior of a Dirac particle within a spacetime background characterized by a specific topological defect, it is necessary to employ the spinor covariant derivative $\nabla _{\mu }$ instead of the partial derivative $\partial _{\mu }$. This allows us to obtain the components of the spinor covariant derivative as $\nabla _{\mu }=\partial _{\mu }+\Gamma _{\mu }\left( x\right)$, where the spinorial connection $\Gamma _{\mu }\left( x\right) =\omega _{\mu ab}\left[ \gamma ^{a},\gamma ^{b}\right] /8$ \cite{7,8,parker,yepez2011einstein} in the local reference frame, correspond to the standard Dirac equation applied to the relativistic quantum description of spin $1/2$--particles in Minkowski background
\begin{equation}
\gamma ^{0}=\hat{\beta}=%
\begin{pmatrix}
\mathrm{I} & 0 \\ 
0 & \mathrm{I}%
\end{pmatrix}%
,\left. {}\right. \gamma ^{i}=\hat{\beta}\hat{\alpha}^{i}=%
\begin{pmatrix}
0 & \sigma ^{i} \\ 
-\sigma ^{i} & 0%
\end{pmatrix}%
,\left. {}\right. 
\end{equation}
and
\begin{equation}
\Sigma ^{i}=%
\begin{pmatrix}
\sigma ^{i} & 0 \\ 
0 & \sigma ^{i}%
\end{pmatrix}%
,  \label{eq:2.5}
\end{equation}%
where, $\Sigma^i$ represents the spin vector, and $\sigma^i$ denotes the Pauli matrices. They follow the standard anti--commutation relations $\{\sigma^i,\sigma^j\}=2\delta_{ij}\mathrm{I}$, where $\delta_{ij}$ is the Kronecker delta and $\mathrm{I}$ is the $2\times 2$ identity matrix. We can also use the notation $\sigma^0=\mathrm{I}$. It is important to note that the indices $i$ and $j$, ranging from 1 to 3, correspond to the spatial components of the local reference frame.

In the context of spinor theory in curved spacetime, we must determine the generalized Dirac matrices using the relation $\gamma ^{\mu}=e_{\left. {}\right. a}^{\mu }\gamma ^{a}$. This allows us to obtain the non--zero components of the spinorial connection. In addition to them, we also need to determine the non--zero components of the connection 1--form $\omega _{\left. {}\right. \mu \left. {}\right. b}^{a}\left( x\right)$. To achieve this, we employ the Cartan structure equations, $\mathrm{d}\hat{\theta}^{a}+\omega_{\left. {}\right. b}^{a}\wedge \hat{\theta}^{b}=0$, under the assumption of vanishing torsion. Here, $\omega_{\left. {}\right. b}^{a}=\omega_{\left. {}\right. \mu \left. {}\right. b}^{a}\left( x\right) dx^{\mu}$, where $\omega_{\left. {}\right. \mu \left. {}\right. b}^{a}\left( x\right)$ are the components of the connection 1--form. By solving the corresponding Cartan structure equations, we obtain the following results:
\begin{equation}
\omega _{\left. {}\right. \varphi \left. {}\right. 1}^{2}=\omega _{\left. {}\right. \varphi \left. {}\right. 2}^{1}=1. \label{eq:2.6}
\end{equation}
Consequently, we find that $\Gamma _{\varphi }=-\mathrm{i}\Sigma ^{3}/2$.


\section{Action of a Dirac spinor field}

In this section, we will examine the behavior of a fermion possessing an EDM when it interacts with an external electric field. Furthermore, we will take into account the gravitational effects that arise from the curvature of spacetime caused by the aforementioned topological defect. The action describing this interaction scenario can be formulated as follows:%
\begin{equation}
\mathcal{S}\left[ \psi \right] =\int \mathrm{d}^{4}x\left. {}\right. \sqrt{%
\left\vert g\right\vert }\mathcal{L},  \label{eq:3.1}
\end{equation}%
where $g=\det \left( g_{\mu \nu }\right) $ represents the determinant of the metric tensor, and the Lagrangian density is given by:%
\begin{equation}
\mathcal{L}=\frac{\mathrm{i}}{2}\bar{\psi}\gamma ^{\mu }\overleftrightarrow{%
\nabla }_{\mu }\psi -m_{\psi }\bar{\psi}\psi .  \label{eq:3.2}
\end{equation}%

The Lagrangian presented here serves as a framework to describe a minimal interaction. It is important to highlight that it incorporates a nonminimal coupling aspect.%
\begin{equation}
\mathrm{i}\gamma ^{\mu }\nabla _{\mu }=\mathrm{i}\gamma ^{\mu }\left(
\partial _{\mu }+\Gamma _{\mu }\right) +\frac{\tilde{d}}{2}\Sigma ^{\mu \nu }%
\tilde{F}_{\mu \nu }.  \label{eq:3.3}
\end{equation}%
In this context, $\tilde{F}_{\mu \nu }$ represents the dual field strength tensor, which specifically characterizes the interaction between the fermion and an external electromagnetic field. In fact, the interaction of the electric dipole moment with
	the electromagnetic field is described by the tensor $\tilde{F}_{\mu \nu }=(-\mathbf{B},-\mathbf{E})$ dual with respect to	the electromagnetic field one ${F}_{\mu \nu }=(-\mathbf{E},\mathbf{B})$ \cite{silenko2005quantum1}.
It is worth noting an anomalous magnetic moment could lead to a much stronger interaction, potentially overshadowing the effect of the electric dipole moment (EDM) due to the latter's comparatively smaller magnitude. However, our focus is solely on the dynamics generated by the EDM on the particle's propagation, ensuring it is not overshadowed by any other effects.

Recent studies have emphasized the importance of isolating the EDM effects in various physical scenarios to better understand the fundamental symmetries of nature. For instance, recent research on the electron EDM experiment using ${\mathrm{HfF}}^{+}$ ions has shown precise control and minimization of systematic uncertainties to enhance the detection sensitivity of the EDM effect \cite{caldwell2023systematic}. Additionally, a new bound on the electron’s EDM was achieved through sophisticated experimental setups designed to reduce magnetic field interference and other potential systematic errors \cite{roussy2023improved}. Another recent study involving $^{171}\mathrm{Yb}$ atoms in an optical dipole trap has provided insights into effective methods of measuring EDM while mitigating external magnetic influences \cite{zheng2022measurement}. The EDM of the neutral Dirac particle is encoded by the coupling constant $\tilde{d}\geq 0$. Furthermore, the components of $\tilde{F}_{\mu \nu }$ are denoted as follows: $%
F_{0i}=-F_{i0}=-E_{i}$ and $F_{ij}=\epsilon _{ijk}B_{k}$ such that the electric and magnetic fields can be expressed as: $\mathbf{E}=\left(
E_{1},E_{2},E_{3}\right) ^{T}$ and $\mathbf{B}=\left(
B_{1},B_{2},B_{3}\right) ^{T}$, respectively. More so, the spin
tensor reads $\sum^{\mu \nu }=\mathrm{i}\left[ \gamma ^{\mu },\gamma
^{\nu }\right] /2$, obeying such a relation $\sum^{\mu \nu }-\sum^{\nu \mu }=%
\mathrm{i}\left[ \gamma ^{\mu },\gamma ^{\nu }\right] $. According to Ref.\cite{silenko2005quantum1}, the anomalous magnetic and electric dipole moments have the lagrangian ${\cal L}_{AMM}	=\frac{\mu^{\prime}}{2}\sigma^{\mu \nu} F_{\mu \nu}$ and ${\cal L}_{EDM}	=-i \frac{d}{2}\sigma^{\mu \nu}\,\gamma^{5} \,F_{\mu \nu}$ respectively. Where $\mu^{\prime}$ is the anomalous particle magnetic moment and $d$ is the Electric Dipole Moment (EDM) of the particle. In the current manuscript $\tilde{d}$ is the Permanent Electric Dipole Moment (PEDM)  of the particle.

Eq. (\ref{eq:3.3}), which encompasses both the spinor covariant derivative due to the topological defect in the background spacetime and an additional term representing the interaction, can be expanded in terms of the components of the electromagnetic field as follows:%
\begin{eqnarray}
\mathrm{i}\gamma ^{\mu }\nabla _{\mu } &=&\mathrm{i}\gamma ^{t}\partial _{t}+%
\mathrm{i}\gamma ^{r}\partial _{r}+\mathrm{i}\gamma ^{\varphi }\left(
\partial _{\varphi }+\Gamma _{\varphi }\right) -\mathrm{i}\tilde{d}\hat{\alpha}^{1}B_{r}+\mathrm{i}\frac{\mu^{\prime} }{r}\hat{%
\alpha}^{2}\left( -B_{\varphi }+\chi B_{r}\right)   \notag \\
&& -\mathrm{i}\tilde{d}\hat{%
\alpha}^{3}B_{z} -\frac{\tilde{d}}{r}\Sigma ^{1}\left( E_{r}+\chi E_{\varphi }\right)  -%
\tilde{d}\Sigma ^{2}E_{\varphi } -\frac{\tilde{d}}{r}\Sigma ^{3}E_{z} +\mathrm{i}\gamma
^{z}\partial _{z}.
\label{eq:3.4}
\end{eqnarray}%

When there is no magnetic field present (i.e., $\boldsymbol{B}=0$), we can derive the complete generalized Dirac equation from the action $\mathcal{S}\left[ \psi \right] $ using the conventional method. The explicit form of the equation is as follows:%
\begin{gather}
\left[ \mathrm{i}\gamma ^{0}\partial _{t}+\mathrm{i}\gamma ^{1}\left(
\partial _{r}+\frac{1}{2r}\right) +\mathrm{i}\gamma ^{2}\frac{1}{r}\left(
\partial _{\varphi }-\chi \partial _{r}\right) +\mathrm{i}\gamma
^{3}\partial _{3}\right.   \notag \\
\left. -\frac{\tilde{d}}{r}\Sigma ^{1}\left( E_{r}+\chi E_{\varphi }\right) -%
\tilde{d}\Sigma ^{2}E_{\varphi }-\frac{\tilde{d}}{r}\Sigma ^{3}E_{z}-m_{\psi
}\right] \psi \left( t,\boldsymbol{r}\right) =0.  \label{eq:3.5}
\end{gather}%
Moving forward, we consider an electric field given by $\boldsymbol{E}=E_{0}r\hat{z}$, where $E_{0}\in \mathbb{R}$ represents a unidirectional electric field along the positive or negative $z$--direction. The field strength increases linearly with radial distance. To describe this scenario, we adopt the following ansatz:
\begin{equation}
\psi \left( t,\boldsymbol{r}\right) =\exp \left[ -\mathrm{i}\mathcal{E}t+%
\mathrm{i}\left( l+\frac{1}{2}\right) \varphi +\mathrm{i}kz\right] \psi
\left( r\right) ,  \label{eq:3.6}
\end{equation}%
and upon solving the complete generalized Dirac equation, we derive a set of four coupled equations governing the behavior of the four components of the radial Dirac spinor $\psi \left( r\right)$. These equations can be expressed as
\begin{subequations}
\label{eq:3.7}
\begin{align}
\left( \mathcal{E}-m_{\psi }-\tilde{d}E_{0}\right) \psi _{1}\left( r\right)
-k\psi _{3}\left( r\right) +\left[ \left( \mathrm{i}-\frac{\chi }{r}\right)
\partial _{r}+\frac{\mathrm{i}}{r}\left( l+1\right) \right] \psi _{4}\left(
r\right) & =0,  \label{eq:3.7a} \\
\left( \mathcal{E}-m_{\psi }+\tilde{d}E_{0}\right) \psi _{2}\left( r\right) +%
\left[ \left( \mathrm{i}+\frac{\chi }{r}\right) \partial _{r}-\frac{\mathrm{i%
}l}{r}\right] \psi _{3}\left( r\right) +k\psi _{4}\left( r\right) & =0,
\label{eq:3.7b} \\
k\psi _{1}\left( r\right) -\left[ \left( \mathrm{i}-\frac{\chi }{r}\right)
\partial _{r}+\frac{\mathrm{i}}{r}\left( l+1\right) \right] \psi _{2}\left(
r\right) -\left( \mathcal{E}+m_{\psi }+\tilde{d}E_{0}\right) \psi _{3}\left(
r\right) & =0,  \label{eq:3.7c} \\
-\left[ \left( \mathrm{i}+\frac{\chi }{r}\right) \partial _{r}-\frac{\mathrm{%
i}l}{r}\right] \psi _{1}\left( r\right) -k\psi _{2}\left( r\right) -\left( 
\mathcal{E}+m_{\psi }-\tilde{d}E_{0}\right) \psi _{4}\left( r\right) & =0.
\label{eq:3.7d}
\end{align}%

In the above equation, the parameter $k\in \mathbb{R}$ represents the eigenvalue of the $z$-component of the linear momentum, given by $\hat{p}_{z}=-\mathrm{i}p_{z}$, while $l\in \mathbb{Z}$ corresponds to the eigenvalue of the angular momentum operator $\hat{L}_{z}=-\mathrm{i}\partial_{\varphi}$. Additionally, $\mathcal{E}$ represents the energy eigenvalue of the Dirac particle, which is the eigenvalue of $\mathrm{i}\partial_{t}$. 

In the following section, we will decouple the set of Eqs. (\ref{eq:3.7}) by deriving a second--order wave equation. This solution will represent the four radial spinor components.


\section{Exact solutions for a Dirac particle}

To decouple the Eqs. (\ref{eq:3.7}), we begin by introducing the ansatz $\psi_1(r) = \xi_1\psi_3(r)$ and $\psi_2(r) = \xi_2\psi_4(r)$, where $\xi_1$ and $\xi_2$ are constants. This choice leads to two conditions, namely \cite{zare2023neutral}
\end{subequations}
\begin{equation}
\xi _{1}=\frac{\mathcal{E}}{k}-\sqrt{\frac{\mathcal{E}^{2}}{k^{2}}-1},\left.
{}\right. \xi _{2}=-\frac{\mathcal{E}}{k}-\sqrt{\frac{\mathcal{E}^{2}}{k^{2}}%
-1}.  \label{eq:4.2}
\end{equation}%
By substituting this outcome into Eqs. (\ref{eq:3.7}), we derive an additional relation linking $\psi_3(r)$ and $\psi_4(r)$%
\begin{equation}
\psi _{4}\left( r\right) =\frac{\left( \mathrm{i}+\frac{\chi }{r}\right)
\partial _{r}-\frac{\mathrm{i}l}{r}}{\left( m_{\psi }-\tilde{d}E_{0}-%
\mathcal{E}\right) \xi _{2}-k}\psi _{3}\left( r\right) .  \label{eq:4.3}
\end{equation}%
Subsequently, we derive a second--order wave equation that governs the behavior of $\psi_3(r)$%
\begin{equation}
\left( 1+\frac{\chi ^{2}}{r^{2}}\right) \frac{\mathrm{d}^{2}\psi _{3}\left( r\right) 
}{\mathrm{d}r^{2}}+\left( \frac{1}{r}-\frac{\chi ^{2}}{r^{3}}-\mathrm{i}\frac{2\chi l%
}{r^{2}}\right) \frac{\mathrm{d}\psi _{3}\left( r\right) }{\mathrm{d}r}-\frac{l^{2}}{r^{2}}%
\psi _{3}\left( r\right) +\mathrm{i}\frac{\chi l}{r^{3}}\psi _{3}\left(
r\right) +\kappa ^{2}\psi _{3}\left( r\right) =0,  \label{eq:4.4}
\end{equation}%
where $\kappa $ assumes the form%
\begin{equation}
\kappa ^{2}=\left( \sqrt{E^{2}+k^{2}}+\tilde{d}E_{0}\right) ^{2}-m_{\psi
}^{2}.  \label{eq:4.5}
\end{equation}%
Prior to delving into the exact solution of this equation, it is pertinent to provide a brief discussion regarding the behavior of $\psi_3(r)$ in the vicinity of small $r$.

When considering small values of $r$, if we assume that the solution of Eq. (\ref{eq:4.4}) follows a power--law behavior of the form $\psi_3(r) \sim r^n$, two distinct solutions arise: $n=2$ and $n=0$. The first solution, $\psi_3(r) \sim r^2$, implies a vanishing wave function at the origin, corresponding to Dirichlet boundary conditions at $r=0$. On the other hand, for the second solution, we extend the ansatz to $\psi_3(r) \sim c + r^m$, where $c$ is an arbitrary constant and $m>0$. This leads to $m=2$, resulting in the boundary condition $\psi_3'(r) = 0$, which corresponds to Neumann boundary conditions at the origin. Both solutions are valid only for $\chi \neq 0$. Notably, $r=0$ is a singular point due to the non--trivial topology described by the line element (\ref{eq:2.1}). While the Dirichlet condition may seem like the natural choice, we will explore both options in our analysis. It is worth mentioning that in the limit of flat space ($\chi=0$), the usual result is obtained: $\psi_3(r) \sim r^{|l|}$. With these considerations, we proceed to find the exact solution of Eq. (\ref{eq:4.4}) by introducing the following ansatz \cite{zare2023neutral}:
\begin{equation}
\psi _{3}\left( r\right) =\exp \left[ \mathrm{i}l\arctan \left( \frac{r}{%
\chi }\right) \right] \widetilde{\mathcal{K}}\left( r\right) ,
\end{equation}%
and equation (\ref{eq:4.4}) yields a wave equation in the form of a Schrödinger--type equation, given by%
\begin{equation}
\left( 1+\frac{\chi ^{2}}{r^{2}}\right) \frac{\mathrm{d}^{2}\widetilde{\mathcal{K}}%
\left( r\right) }{\mathrm{d}r^{2}}+\frac{1}{r}\left( 1-\frac{\chi ^{2}}{r^{2}}\right) 
\frac{\mathrm{d}\widetilde{\mathcal{K}}\left( r\right) }{\mathrm{d}r}-\frac{l^{2}}{r^{2}+\chi
^{2}}\widetilde{\mathcal{K}}\left( r\right) +\kappa ^{2}\widetilde{\mathcal{K%
}}\left( r\right) =0.  \label{eq:4.6}
\end{equation}%
In the next step, we perform a change of variables by introducing a new radial variable $\rho = \sqrt{\chi^2 + r^2}$ with $\rho \geq \chi$, and define $\mathcal{K}(\rho) = \widetilde{\mathcal{K}}(\sqrt{\rho^2 - \chi^2})$. By making this substitution, we obtain the following expression:%
\begin{equation}
\frac{\mathrm{d}^{2}\mathcal{K}\left( r\right) }{\mathrm{d}\rho ^{2}}+\frac{1}{\rho }\frac{\mathrm{d}%
\mathcal{K}\left( r\right) }{\mathrm{d}\rho }+\frac{1}{\rho ^{2}}\left( \kappa
^{2}\rho ^{2}-l^{2}\right) \mathcal{K}\left( r\right) =0.  \label{eq:4.7}
\end{equation}%

Remarkably, Eq. (\ref{eq:4.7}) takes the form of the radial Schr\"{o}dinger equation describing a free particle with mass 1/2 and energy $\varepsilon = \kappa^2$ moving in a two--dimensional plane. Here, $\kappa > 0$ can be interpreted as the wave number for the radially outgoing and incoming waves. However, it is important to note that the radial variable $\rho$ is constrained to satisfy $\rho \geq \chi$, which means that the two--dimensional plane has a hole at the origin with a radius of $\chi$. Consequently, we need to impose boundary conditions at the boundary of this hole. Based on our previous analysis of small $r$, we can consider both Dirichlet and Neumann conditions at $\rho = \chi$.

Recognizing that Eq. (\ref{eq:4.7}) corresponds to the well--known Bessel differential equation, we can express its general solution in terms of two linearly independent Bessel functions. In this case, we choose the Hankel functions of the first and second kind, denoted as $H_{l}^{\left( 1\right)}\left(\kappa \rho\right)$ and $H_{l}^{\left( 2\right)}\left(\kappa \rho\right)$, respectively. These functions exhibit the asymptotic behavior $H_{l}^{\left( 1,2\right)}\left(\kappa \rho\right) \sim \exp\left(\pm \kappa \rho\right) \sim \exp\left(\pm \kappa r\right)$ for large values of $\rho \sim r$, representing outgoing and incoming radial waves with the expected wave number $\kappa$. The exact solution of Eq. (\ref{eq:4.7}) is given by:
\begin{equation}
\mathcal{K}\left( r\right) =\mathcal{A}H_{l}^{\left( 1\right) }\left( \kappa
\rho \right) +\mathcal{B}H_{l}^{\left( 2\right) }\left( \kappa \rho \right) .
\label{eq:4.8}
\end{equation}%
From Dirichlet and Neumann boundary conditions, we have%
\begin{equation}
\mathcal{B}^{D}=-\mathcal{A}^{D}\frac{H_{l}^{\left( 2\right) }\left( \kappa
\chi \right) }{H_{l}^{\left( 1\right) }\left( \kappa \chi \right) },\left.
{}\right. \mathcal{B}^{N}=-\mathcal{A}^{N}\frac{H_{l}^{\left( 2\right)
\prime }\left( \kappa \chi \right) }{H_{l}^{\left( 1\right) \prime }\left(
\kappa \chi \right) },  \label{eq:4.9}
\end{equation}%
and the energy eigenvalues read%
\begin{equation}
\mathcal{E}_{k,\kappa }=\pm \sqrt{k^{2}+\left( \sqrt{m_{\psi }^{2}+\kappa
^{2}}-\tilde{d}E_{0}\right) ^{2}},\left. {}\right. k\in 
\mathbb{R}
\left. {}\right. \kappa \geq 0.  \label{eq:4.10}
\end{equation}%

The energy spectrum of a neutral Dirac particle in a background with a topological defect and a Permanent Electric Dipole Moment (PEDM) $\tilde{d}$ coupled to an electric field exhibits a continuous spectrum bounded from below. Similar to the case of a magnetic field pointing towards the positive or negative z--axis, the presence of the PEDM in an electric field either decreases or increases the usual spectral gap of the free Dirac fermion \cite{zare2023neutral}. Remarkably, the deformation parameter $\chi$ does not affect the spectrum itself, but it clearly influences the wave functions (\ref{eq:4.8}) through the parameters (\ref{eq:4.9}) and the $r$--dependent phase $\exp \left[ \mathrm{i}l\arctan \left(
r/\chi \right) \right]$. Another point worthy to be explored is the consequences of Eq. (\ref{eq:4.10}) in the thermodynamic properties of the system. Based on this aspect, recently in the literature, many studies have been accomplished within variety of contexts, namely, loop quantum gravity \cite{aa2022particles}, Einstein--aether theory \cite{aaa2021thermodynamics}, quantum gases \cite{araujo2022does,reis2022quantum}, quantum rings \cite{oliveira2019thermodynamic,araujo2022thermodynamics}, traversable wormholes \cite{aa14}, rainbow gravity \cite{furtado2023thermal}, bouncing universe \cite{petrov2021bouncing2}, graviton \cite{aa2021lorentz}, higher--dimensional operators \cite{araujo2021higher,reis2021thermal,araujo2021thermodynamic}, fermions on a torus \cite{araujo2022fermions} and others \cite{oliveira2020relativistic,oliveira2020thermodynamic,araujo2023gravitational,heidari2023gravitational,aa13,aa2023analysis,aa2024implications}.

It is noteworthy that the system being investigated demonstrates a SUSY structure resembling SUSY quantum mechanics, particularly in the absence of electric interaction. This observation aligns with recent findings in the literature \cite{zare2023neutral}.


The Aharonov--Anandan (AA) geometrical phase is a fundamental concept in quantum mechanics that arises from the geometric properties of the quantum state space \cite{aharonov1987phase}. It provides valuable insights into the underlying geometry of physical systems and has been extensively studied in various contexts. In this regard, understanding the AA phase developed by the spinor is of particular interest. The spinor, as a mathematical representation of a quantum particle with spin, carries essential information about its intrinsic angular momentum and associated symmetries. Investigating the AA phase in spinor systems allows us to explore the intricate interplay between geometric properties, such as curvature, topology, and the quantum dynamics of spin and mojorana neutrinos \cite{e1,e2,e3,e4}. Moreover, recent developments in the literature have unveiled connections between the AA phase and other phenomena, such as topological phases of matter and supersymmetry \cite{zare2023neutral}.

Now, we evaluate the phases the AA geometrical phase
developed by the spinor \cite{17,25}, which is given by:%
\begin{equation}
\Phi _{\mathrm{AA}}=\mathrm{i}\int_{0}^{2\pi }\psi ^{\dag }\left( t,%
\boldsymbol{r}\right) \frac{\partial \psi \left( t,\boldsymbol{r}\right) }{%
\partial \varphi }\mathrm{d}\varphi .
\end{equation}%
For the cyclic motion around the $z$--axis we found that%
\begin{equation}
\Phi _{\mathrm{AA}}=-2\pi \left( l+\frac{1}{2}\right) \left\vert \psi \left(
r\right) \right\vert ^{2}.
\end{equation}
where $l$ is a quantum number characterizing the angular momentum and $\left|\psi(r)\right|^2$ represents the squared magnitude of the spinor wavefunction. This expression reveals that the AA phase depends on both the quantum number $l$ and the spatial distribution of the spinor wavefunction $\psi(r)$. The term $-2\pi \left(l + \frac{1}{2}\right)$ indicates the contribution from the quantum number, while $\left|\psi(r)\right|^2$ represents the spatial dependence of the wavefunction.

The AA phase is a geometric phase that arises from the non-integrability of the system's Hamiltonian and characterizes the accumulation of a phase factor during cyclic evolutions in the parameter space. It has significance in various fields, including quantum mechanics, quantum computation, and quantum information processing. Further analysis and investigation of the AA phase in different systems and scenarios can provide valuable insights into the geometric properties of quantum states and their applications in quantum technologies.


\section{The thermodynamic properties}

The grand partition function, which characterizes the grand canonical ensemble, can be mathematically represented as follows:
\begin{equation}
\Xi =\sum_{N=0}^{\infty }\exp \left( \beta \mu N\right) \mathcal{Z}\left[
N_{n,s}\right] ,  \label{eq:GarndPartition-function}
\end{equation}%
where $\mathcal{Z}\left[ N_{n,s}\right]$ is the canonical partition function. In the context of fermions, the occupation number of each quantum state is subject to a constraint where it can only be either 0 or 1. Consequently, the energy of a particular quantum state is contingent upon its corresponding occupation number
\begin{equation}
E\left\{ N_{n,s}\right\} =\sum_{\left\{ n,s\right\} }N_{n,s}E_{n,s}
\end{equation}%
here,
\begin{equation}
\sum_{\left\{ n,s\right\} }N_{n,s}=N.
\end{equation}%
Therefore, the partition function becomes:
\begin{equation}
\mathcal{Z}\left[ N_{n}\right] =\sum_{\left\{ N_{n,s}\right\} }\exp \left[
-\beta \sum_{n=0}^{\infty }\sum_{s=1}^{2}N_{n,s}E_{n,s}\right] .
\end{equation}%
In addition, the grand partition function takes the form%
\begin{equation}
\Xi =\sum_{N=0}^{\infty }\exp \left( \beta \mu N\right) \sum_{\left\{
N_{n,s}\right\} }\exp \left[ -\beta \sum_{n=0}^{\infty
}\sum_{s=1}^{2}N_{n,s}E_{n,s}\right] ,
\end{equation}%
which can be rewritten as
\begin{equation}
\Xi =\prod_{n}\prod_{s}\left\{ \sum_{N_{n,s}=0}^{1}\exp \left[ -\beta
N_{n,s}\left( E_{n,s}-\mu \right) \right] \right\} .
\end{equation}%
By performing the sum over the two possible occupation numbers, we obtain:
\begin{equation}
\Xi =\prod_{n}\prod_{s}\left\{ 1+\exp \left[ -\beta\left(
E_{n,s}-\mu \right) \right] \right\} .
\end{equation}
To connect with thermodynamics, we introduce the grand potential:
\begin{equation}
\Phi =-\frac{1}{\beta }\ln \Xi .
\end{equation}%
Replacing $\Xi$ in above expression, we get%
\begin{equation}
\Phi =-\frac{1}{\beta }\sum_{n=0}^{\infty }\sum_{s=1}^{2}\ln \left\{ 1+\exp %
\left[ -\beta \left( E_{n,s}-\mu \right) \right] \right\} .
\label{eq:Gand-potential}
\end{equation}%

While it is possible to employ the Euler--Maclaurin formula for calculations, the resulting outcomes tend to be extensive and challenging to interpret in the context of their physical implications. To ensure clarity in our analysis, we will instead rely on numerical calculations. Additionally, we will utilize the grand potential to compute various thermodynamic quantities of significance, such as the mean particle number, internal energy, entropy, and heat capacity
\begin{equation}
\mathcal{N}=-\frac{\partial \Phi }{\partial \mu },\phantom{a}\mathcal{U}%
=-T^{2}\frac{\partial }{\partial T}\left( \frac{\Phi }{T}\right) ,\phantom{a}%
S=-\frac{\partial \Phi }{\partial T},\phantom{a}C=T\frac{\partial S}{%
\partial T}.
\end{equation}

In order to compute the thermodynamic quantities using both methods, it is necessary to evaluate the summation given by Eq. $\left( \ref{eq:Gand-potential}\right)$. Unfortunately, this sum cannot be computed analytically. However, we can resort to numerical analyses, which will provide valuable insights into the behavior of the thermodynamic quantities. In our analysis, we adopt the values $\mathrm{m}=0.511~\mathrm{MeV}$ for the electron mass and $\mathrm{\mu}=0.1~\mathrm{eV}$ for the chemical potential. Our primary focus lies in examining the variations of these quantities as the electric field undergoes changes. To aid in the interpretation of the results, we introduce the parameter $\xi=\tilde{d}E_{0}$, which governs this variation.

\begin{figure}[tbh]
\centering
  \subfloat[Number of particles]{\includegraphics[width=8cm,height=5cm]{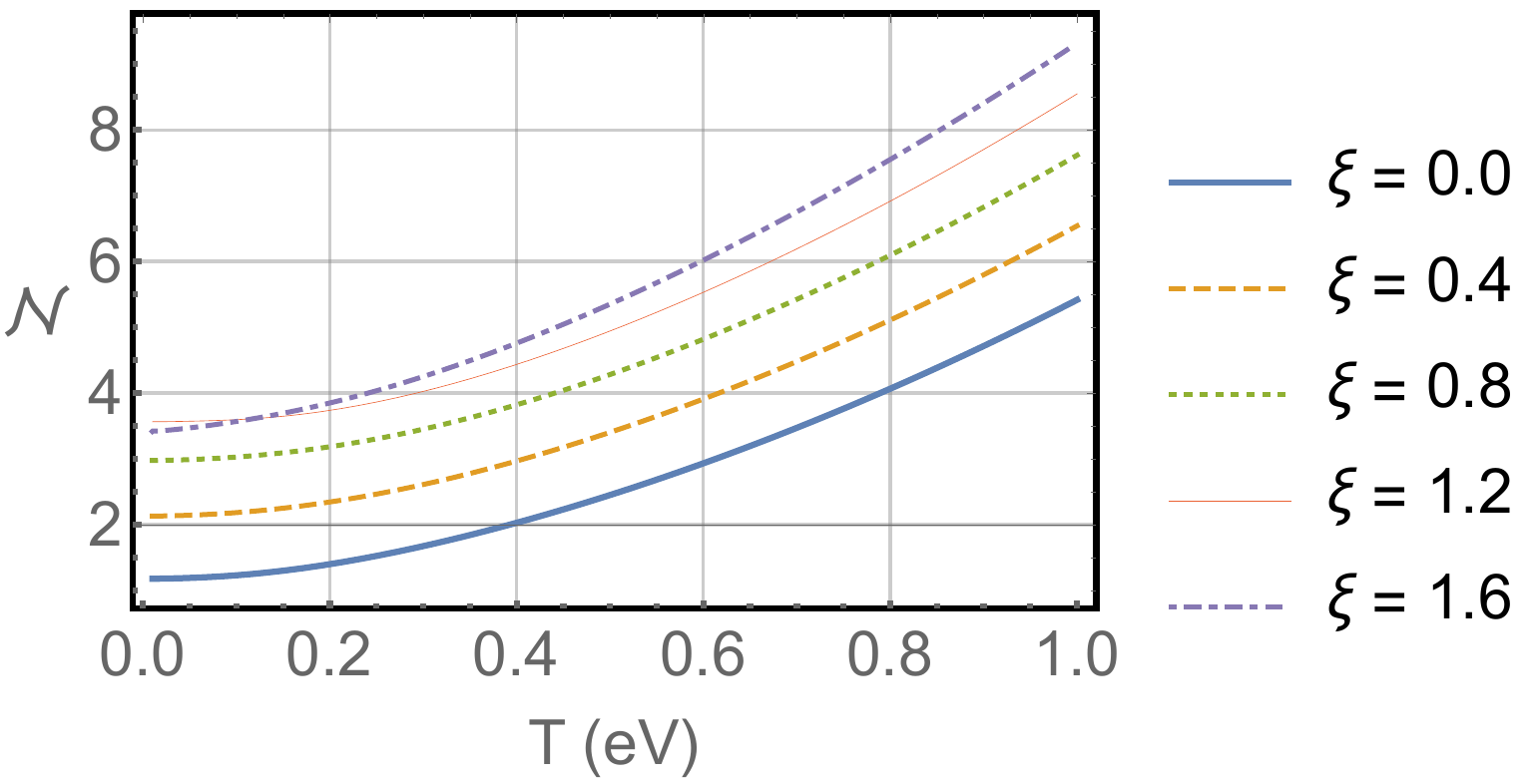}
  \label{fig:GPCase-2-lowTN}}
  \subfloat[Internal energy]{\includegraphics[width=8cm,height=5cm]{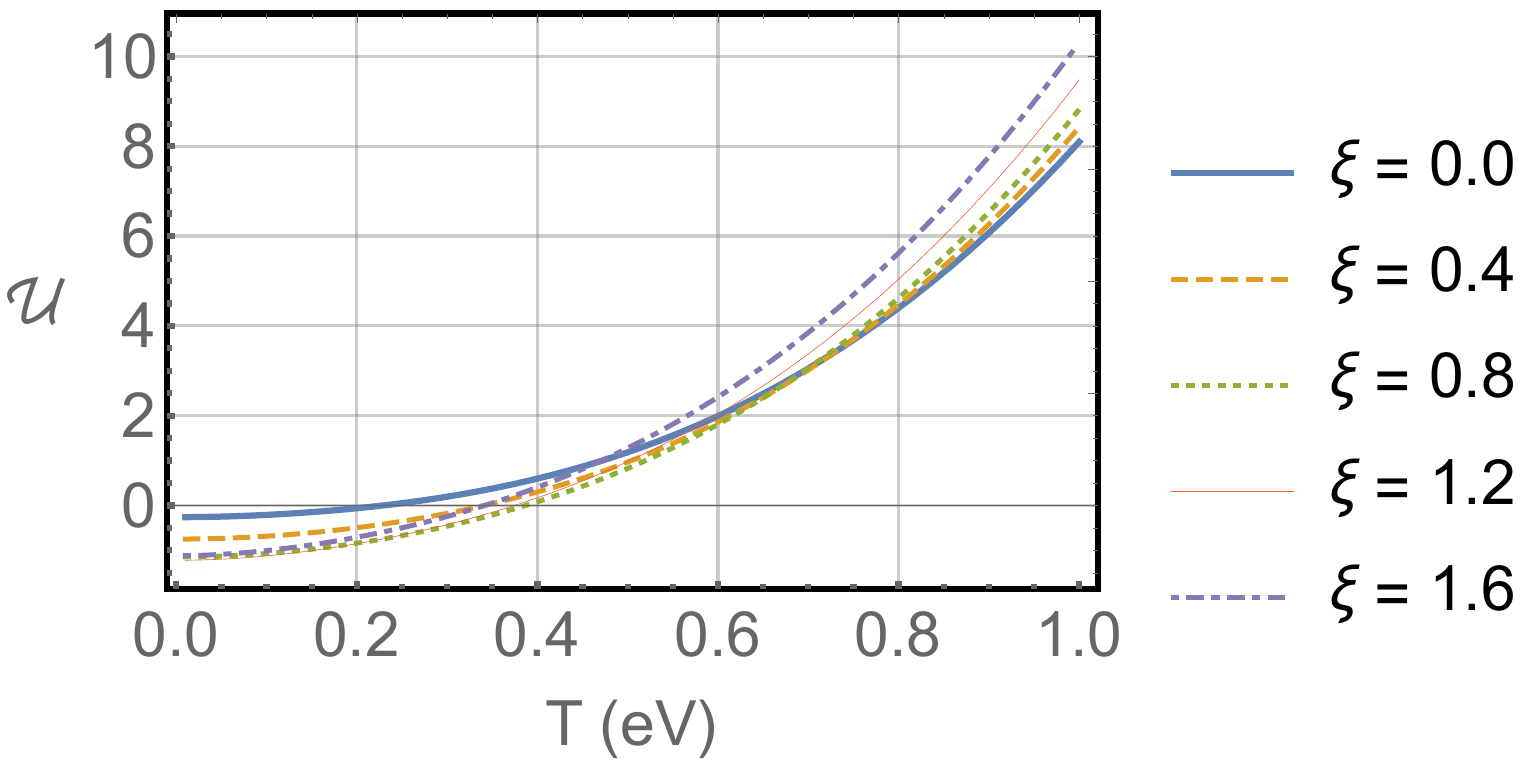}
  \label{fig:GPCase-2-lowTU}}\\
  \subfloat[Entropy]{\includegraphics[width=8cm,height=5cm]{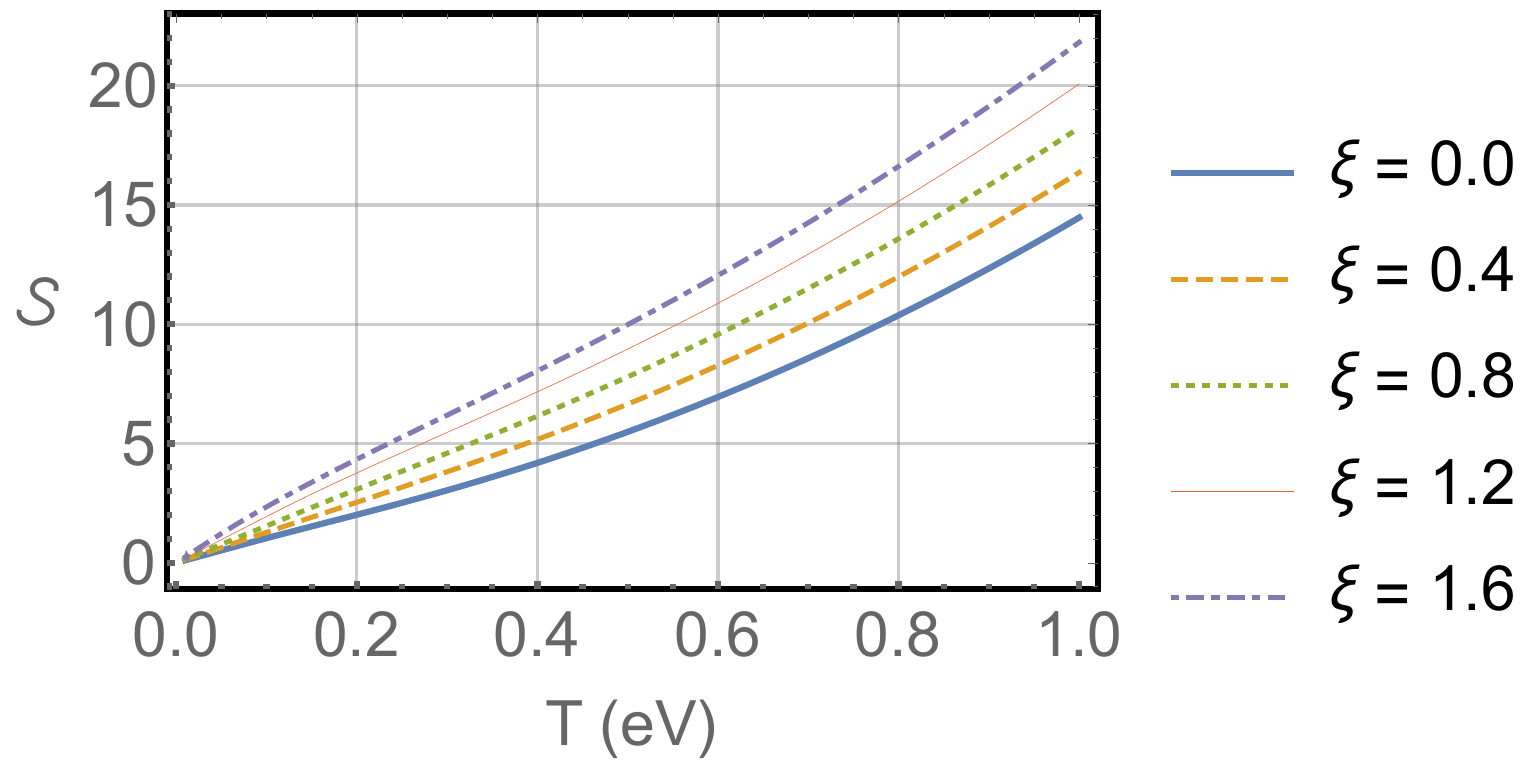}
  \label{fig:GPCase-2-lowTS}}
  \subfloat[Heat capacity]{\includegraphics[width=8cm,height=5cm]{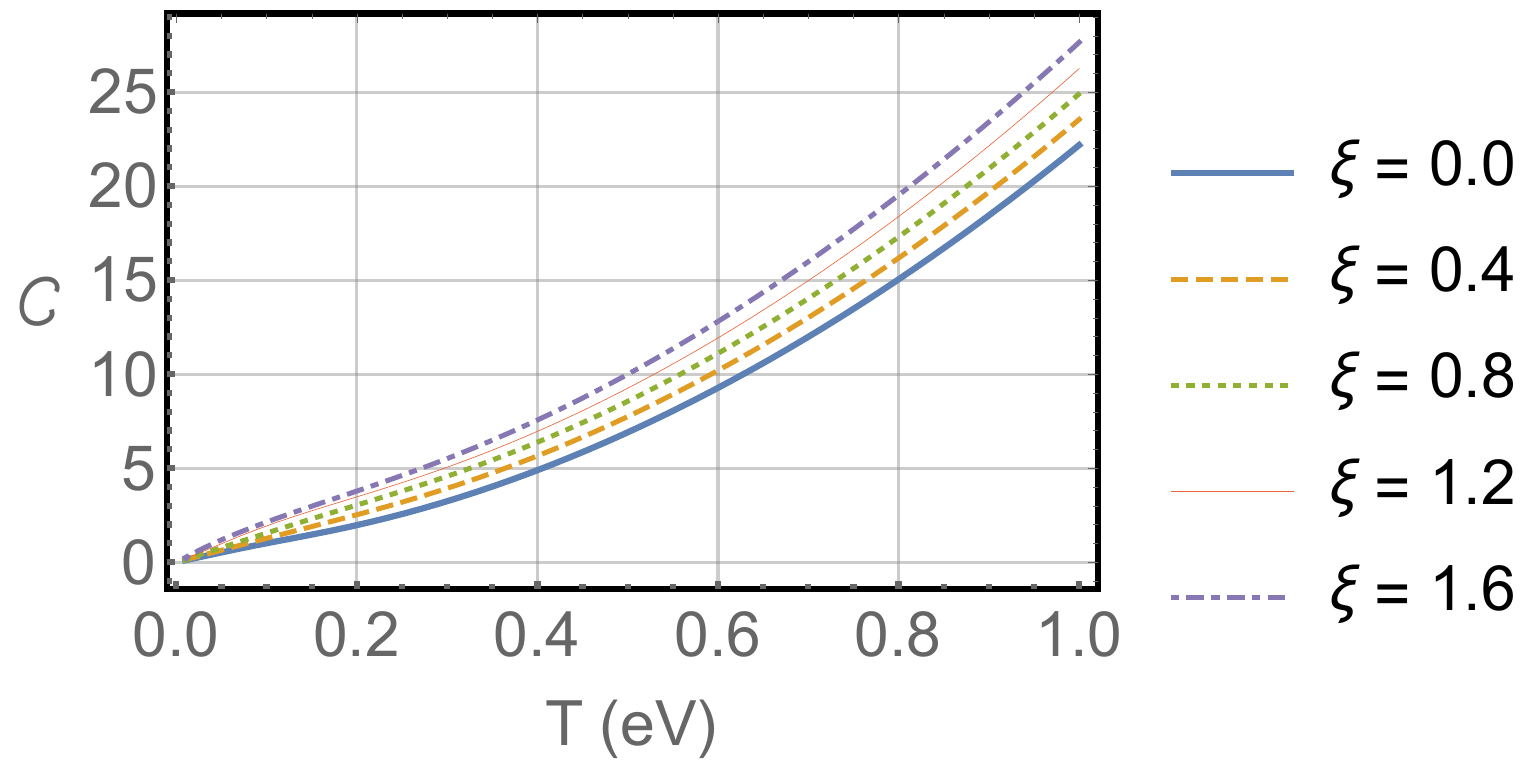}
  \label{fig:GPCase-2-lowTC}}
 \caption{Number of particles, internal energy, entropy and heat capacity are displayed for different values of $\xi$.}
\end{figure}

Let us begin by examining the particle number, as depicted in Fig. \ref{fig:GPCase-2-lowTN}. It is evident that the fluctuations in this quantity amplifies with escalating temperature and electric field intensity. Moving on to the internal energy, as illustrated in Fig. \ref{fig:GPCase-2-lowTU}, we observe two distinct regimes. For temperatures below $T = 0.5~\mathrm{eV}$, the internal energy decreases as the electric field increases. However, for temperatures surpassing $T = 0.5~\mathrm{eV}$, the internal energy demonstrates an opposing trend, indicating an increase with higher electric field values. In terms of entropy, displayed in Fig. \ref{fig:GPCase-2-lowTS}, we see an ascending pattern as the temperature rises.

More so, it is worth noting that the entropy exhibits an increase with the elevation of the electric field. Lastly, let us explore the behavior of the heat capacity in relation to temperature, as illustrated in Fig. \ref{fig:GPCase-2-lowTC}. It is evident that the heat capacity is responsive to variations in the electric field. Furthermore, the heat capacity displays an ascending trend as the temperature increases. It is important to emphasize that this analysis is based on the specific case discussed, involving the grand canonical potential, which allows for an examination of particle number fluctuations and other related phenomena.

\begin{figure}[tbh]
\centering
  \subfloat[Permittivity]{\includegraphics[width=8cm,height=5cm]{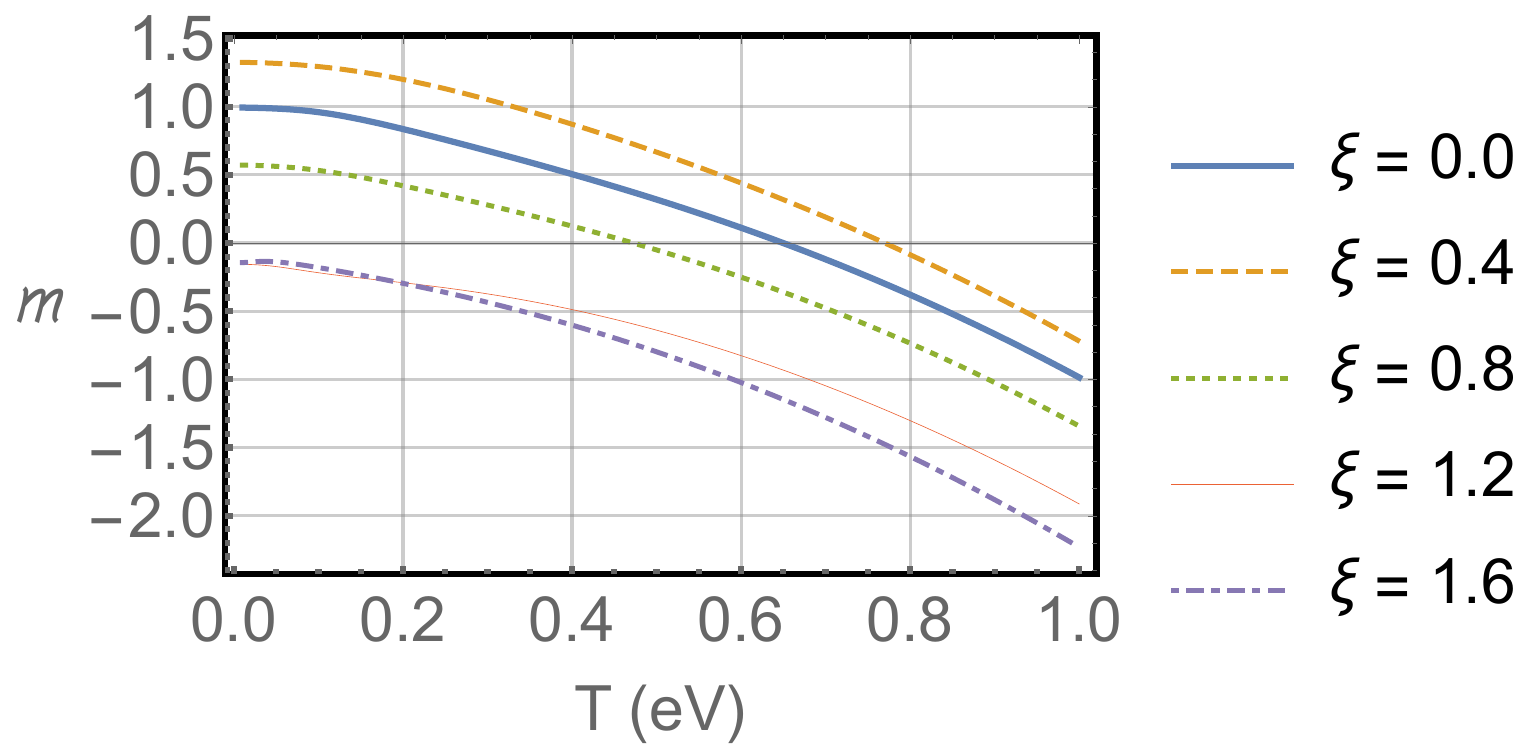}
  \label{fig:GPCase-2-lowTm}}
  \subfloat[Susceptibility]{\includegraphics[width=8cm,height=5cm]{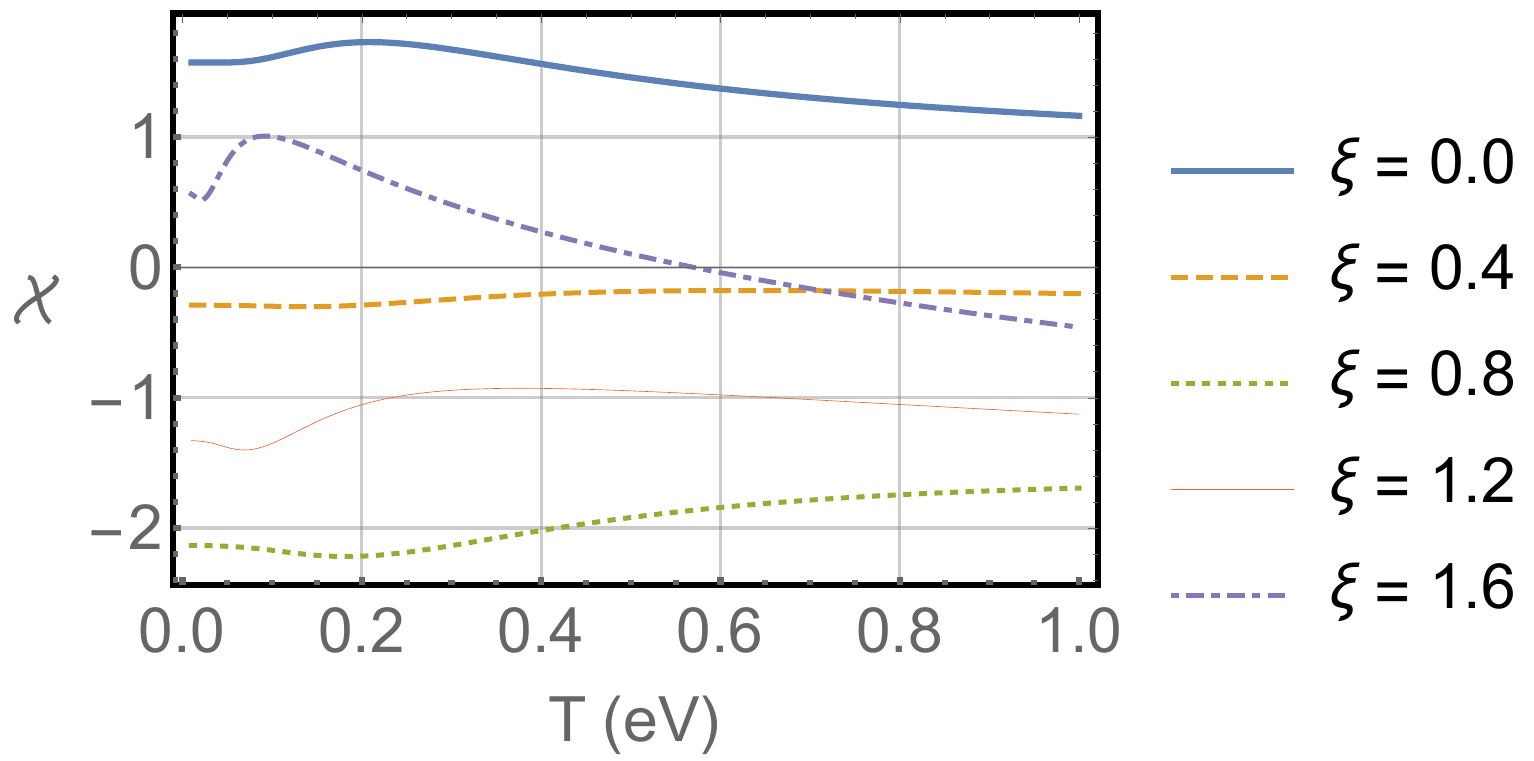}
  \label{fig:GPCase-2-lowTx}}
  \caption{Permittivity and susceptibility are shown for different values of $\xi$}
\end{figure}

Finally, we can analyze the permittivity and the susceptibility of the system. Those quantities are given, respectively, by \cite{pathria2016statistical}
\begin{equation}
m=-\frac{\partial \mathcal{U}}{\partial E},\phantom{ss}\chi=\frac{\partial m}{\partial E}.
\label{eq:Therm-functions2}
\end{equation}%
Permittivity and susceptibility are fundamental properties with profound implications in diverse scientific domains. Permittivity quantifies the capacity of a medium to facilitate the propagation of the electric field, while susceptibility describes the response of the system to it. These properties play crucial roles in comprehending the dynamics of electromagnetic fields, including their interactions within curved spacetimes and gravitational fields.

Fig. \ref{fig:GPCase-2-lowTm} provides insights into the behavior of the permittivity with changing temperature. It is evident that the permittivity decreases as the temperature rises. Moreover, for $T = 0$, it is worth noting that the permittivity is non--zero and depends on the electric field magnitude. 

Intriguingly, the susceptibility, as depicted in Fig. \ref{fig:GPCase-2-lowTx}, exhibits a compelling pattern that is influenced by the external electric field. Notably, at a temperature of $T = 0.1~\mathrm{eV}$ and a parameter value of $\xi = 1.6$, a global maximum is observed. Conversely, at the same temperature but with $\xi = 1.2$, a global minimum is observed. Therefore, we can deduce the existence of a transition region between these values where the system undergoes a shift from a global minimum to a global maximum value for the susceptibility.

The observed transition region in the susceptibility corresponds to a specific range of values for the external electric field. Within this range, as the temperature remains constant, the susceptibility changes from reaching a global minimum to reaching a global maximum value. This transition indicates a significant change in the system's response to the applied electric field.

At lower electric field strengths within this transition region, the susceptibility reaches a global minimum value. This suggests that the system is less responsive to the electric field, exhibiting reduced polarization or alignment of its internal dipoles. The decrease in susceptibility indicates a decreased ability of the system to be influenced by the external electric field.

Conversely, at higher electric field strengths within the transition region, the susceptibility reaches a global maximum value. This implies that the system becomes more responsive to the electric field, showing enhanced polarization or alignment of its internal dipoles. The increase in susceptibility indicates an increased ability of the system to be influenced and polarized by the external electric field.


\section{Conclusion}

In this paper, we began by considering a $(1+3)$--dimensional spacetime that emerges from the presence of a spiral dislocation background. We provided a concise overview of the mathematical formulation of the Dirac spinor within the context of quantum field theory in curved spacetime.

Subsequently, we investigated the relativistic behavior of a spin--half particle with a PEDM interacting with an external electric field in this background. We formulated the action of a Dirac spinor field, incorporating the Lagrangian density of the Dirac spinor field in the background and the Lagrangian density corresponding to the interaction model. By seeking analytical solutions, we established the wave equation derived from the action of the Dirac spinor field in this specific scenario.

We successfully reduced the relativistic wave equations to a radial Schrödinger problem for a free particle moving on a planar surface with a hole of radius determined by the deformation parameter $\chi$. The energy eigenvalues were obtained in closed form, revealing that this system possesses a continuous spectrum akin to that of a free Dirac particle. However, the presence of the PEDM alters the gap between the positive and negative branches. Additionally, we presented the closed--form expressions for the Dirac energy eigenspinors, considering both Dirichlet and Neumann boundary conditions along the $z$--axis, both of which are admissible.

Finally, to validate our findings, we also focused on the following applications: the study of the geometric phase associated with the system and the thermodynamic properties, where a phase transition was verified.

\section*{Acknowledgments}
\hspace{0.5cm}

The authors would like to thank the referees for their useful suggestions that improved the paper. The authors also express their gratitude to FAPEMA, CNPq and CAPES (Brazilian research agencies) for invaluable financial support. In particular, L. Lisboa-Santos is supported by Conselho Nacional de Desenvolvimento Cientifico e Tecnologico CNPQ/PDJ 151424/2022 and A. A. Araújo Filho is supported by Conselho Nacional de Desenvolvimento Cientíıfico e Tecnológico (CNPq) and Fundação de Apoio à Pesquisa do Estado da Paraíba (FAPESQ) -- 200486/2022-5 and 150891/2023-7.

\section{Data Availability Statement}

Data Availability Statement: No Data associated in the manuscript


\bibliographystyle{ieeetr}
\bibliography{main}

\end{document}